\documentclass[prb,twocolumn,superscriptaddress]{revtex4} 
\usepackage{graphicx}
\usepackage{dcolumn}
\usepackage{bm}
\usepackage[latin1]{inputenc}
\usepackage{amsfonts}
\usepackage{amssymb}
\usepackage{amsmath}
\begin{document}
\title{Optical spin control in nanocrystalline magnetic nanoswitches}      

\author{C. Echeverr\'ia-Arrondo}
\affiliation{Departamento de F\'\i sica de Materiales, Facultad de Qu\'\i micas, Centro de F\'isica de Materiales, CSIC-UPV/EHU, E-20018, San Sebasti\'an/Donostia, Spain and Donostia International Physics Center (DIPC), E-20018 San Sebasti\'an/Donostia, Spain}
\affiliation{Departamento de F\'isica, Universidad P\'ublica de Navarra, E-31006, Pamplona, Spain}

\author{J. P\'erez-Conde}
\affiliation{Departamento de F\'isica, Universidad P\'ublica de Navarra, E-31006, Pamplona, Spain}


\author{A. Ayuela}
\affiliation{Departamento de F\'\i sica de Materiales, Facultad de Qu\'\i micas, Centro de F\'isica de Materiales, CSIC-UPV/EHU, E-20018, San Sebasti\'an/Donostia, Spain and Donostia International Physics Center (DIPC), E-20018 San Sebasti\'an/Donostia, Spain}

\begin{abstract}
We investigate the optical properties of (Cd,Mn)Te quantum dots (QDs) by looking at the excitons as a function of the Mn impurities positions and their magnetic alignments. When doped with two Mn impurities, the Mn spins, aligned initially antiparallel in the ground state, have lower energy in the parallel configuration for the optically active spin-up exciton. Hence, the photoexcitation of the QD ground state with antiparallel Mn spins induces one of them to flip and they align  parallel. This suggests that (Cd,Mn)Te QDs are suitable for spin-based operations handled by light. 
\end{abstract}
\maketitle

Light control of magnetic dopants \cite{hanson} in diluted magnetic semiconductor nanocrystals (DMS NCs) is a powerful tool for spin manipulation in spintronics\cite{wolf}.
The  fabricated DMS NCs of II-VI compounds show outstanding magneto-optical properties \cite{bhargava} such as large Zeeman splittings and excitonic magnetic polarons (EMPs). 
The experimental studies based on spectroscopic techniques confirm that (i) Mn impurities can be actually embedded within II-VI QDs \cite{erwin,norris,besombes} and that (ii) stable EMPs can be induced by light, especially in Mn-doped CdTe dots \cite{mackowski,maksimov,gurung,kuroda}.  These results  have motivated some theoretical approaches to EMPs in DMS QDs\cite{bhatta1,bhatta4,xia,bhatta5,rossier,chang,Li} based on the effective mass approximation (EMA). Although such approximation is widely used,  first-principles methods for small nanoparticles give a more realistic description of the electronic density and thus of the electron-hole exchange interaction together with their magneto-optical properties \cite{franceschetti,carlos}.

\begin{figure}[b]
\centering
\includegraphics[scale=0.25]{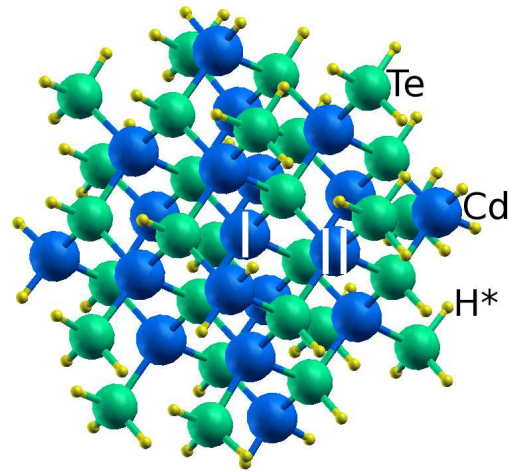}
\caption{\label{fig:fig1}(Color online) Geometry of the studied nanocrystals. The atoms are Cd in dark grey (blue), Te in light grey (green), and pseudohydrogens $\rm{H}^{\ast}$ with small atoms (yellow). The considered cationic sites for Mn are grouped in two sets: ``I'' stands for QD center and ``II'' stands for Cd sites off center.}
\end{figure}

Previous works\cite{chang,Li} dealt with static magnetic and electric fields in single doped Mn quantum dots within the EMA and  using the mean field approach.
However, in this work we investigate the dot luminescence of small CdTe NCs doped with several Mn where the many body effects are dealt within density functional theory. 
The  QDs are quasi-spherical, of $\sim2~$nm in diameter, as shown in Fig.~\ref{fig:fig1}. We consider one and two Mn impurities in different positions and in the later case with different magnetic couplings. For two Mn atoms, the Mn spins are antiparallel in the most stable magnetic configuration, but when an exciton is created they turn parallel and yield the formation of an EMP. This magnetic effect is induced by the hole, which is coupled to the Mn spins and described as an effective exchange mechanism. The photocreated EMP is sufficiently stable in time to permit spin-based operations with (Cd,Mn)Te QDs, which behave under optical excitation as effective magnetic nanoswitches.  


We study NCs made of a central cation (Cd$^{2+}$) and consecutive layers of (Cd,Mn)Te with zinc-blende symmetry, as given in Fig.~\ref{fig:fig1}. The spherical NC measures  about $17$~{\AA} in diameter.  The NCs are saturated with pseudohydrogen atoms (H*), which prevent surface states from appearing in the near-gap spectrum, therefore we simulate Mn-doped colloidal dots \cite{besombes}.
For the NC calculations we use the projector-augmented wave method, as implemented in the VASP code.\cite{kresse2,kresse3} We take into account the \textit{s,p} valence electrons of (Cd,Mn)Te as well as the $d$ electrons of Cd and Mn.
For the exchange and correlation interactions, we use the generalized gradient functional \cite{perdew,freeman}  (GGA+\textit{U}) as in Ref. \onlinecite{carlos}.
We relax the initial QD geometry with CdTe bulk distances until the atomic forces are smaller than 0.02 eV/{\AA}. The cut-off energy for the plane waves is 350 eV.

We investigate the excitons of lowest energy. In our calculations, four types of excitons are possible, namely $|\Downarrow\uparrow\rangle$, $|\Uparrow\downarrow\rangle$, $|\Uparrow\uparrow\rangle$, and $|\Downarrow\downarrow\rangle$,  where $\uparrow$ ($\downarrow$) stands for spin-up (spin-down) electron and $\Uparrow$ ($\Downarrow$) for spin-up (spin-down) hole, following the notation of Ref.~\onlinecite{pumping}. In accordance with the standard selection rules, the excitons $|\Downarrow\uparrow\rangle$ and $|\Uparrow\downarrow\rangle$ are spin allowed or optically active, also known as bright excitons.  Similarly, the excitons $|\Uparrow\uparrow\rangle$ and $|\Downarrow\downarrow\rangle$ are spin forbidden or optically inactive, also known as dark excitons. We note that the orbital selection rule is always fulfilled by bright and dark excitons, as the hole is $p$-type and the electron is $s$-type.

\begin{figure}[!t]
\centering
\includegraphics[scale=0.85]{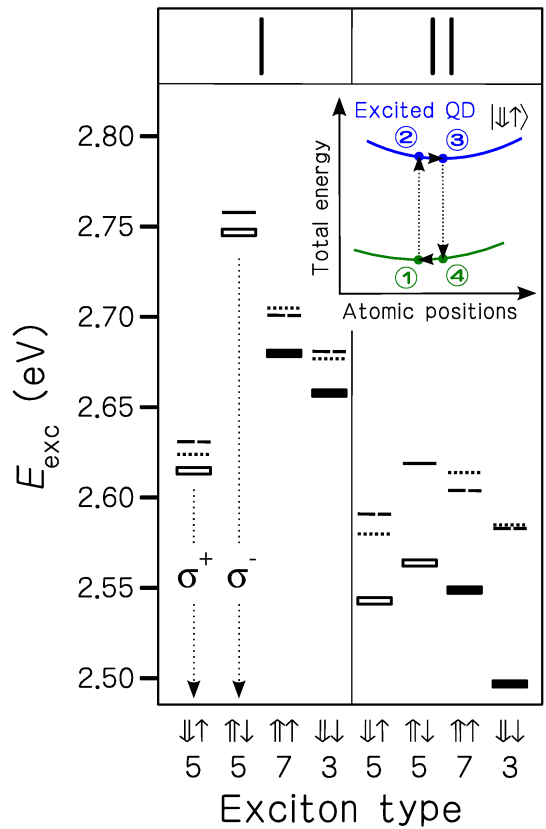}
\caption{\label{fig:fig2}(Color online)  Excitation energies of QDs doped with a single Mn impurity as a function of the exciton type and the impurity position, either I or II. Open symbols are for bright excitons in relaxed QD geometries and closed symbols are for dark excitons; dashed lines stand for vertical excitation energies of unrelaxed QD geometries; and dotted lines stand for fundamental gaps in the Mn-doped NCs. For bright excitons $\mu_{\rm{QD}}=5~\mu_{B}$ and for dark excitons $\mu_{\rm{QD}}$ is either 7~$\mu_{B}$ or 3~$\mu_{B}$.
The inset shows a total energy scheme for QDs doped with a single Mn impurity. Atomic positions are fully relaxed in the adiabatic states (1) and (3). 
We observe that the $\sigma^+$ and $\sigma^-$ splitting depends on the Mn position. Note that for position II, close to the surface, the dark exciton $|\Downarrow\downarrow\rangle$ has the lowest energy.
}
\end{figure}

The excitation energy $E_{\rm{exc}}$ is defined as the difference between the total energy in the QD after excitation $E^{\ast}$ and the total ground-state energy \textit{E}, that is $E_{\rm{exc}}=E^{\ast}-E$. This expression takes into account the whole interaction between electron and hole. The total energies of QDs are schematically shown in the inset of Fig.~\ref{fig:fig2}. They correspond to QDs in excited and unexcited states. The energy difference $E_{\rm{exc}}$ between the states (1) and (2) corresponds to a vertical excitation, and $E_{\rm{exc}}$ between the adiabatic states (1) and (3) corresponds to an excitation between fully relaxed geometries. The calculations of total energies are carried out by fixing (i) occupancies of one-electron states and (ii) total electronic spins ($\mu^{\ast}_{\rm{QD}}$).

Figure~\ref{fig:fig2} shows the excitation energy $E_{\rm{exc}}$ against Mn position and exciton type. When  Mn is located in the NC center (site I), $E_{\rm{exc}}$ is larger than when placed off center (site II).  Moreover, the   splitting $\Delta E$ between $\sigma^+$ and $\sigma^-$ is larger when Mn replaces a Cd cation in the QD center (site I, $\Delta E^{\rm{I}}=$~127 meV) than when a Cd is off center (site II, $\Delta E^{\rm{II}}=28$~meV). These differences follow the fundamental gaps for Mn in the NC positions (dotted lines in Fig.~\ref{fig:fig2}). Thus, from $\Delta E^{\rm{I}}$ we estimate the effective magnetic field induced by the impurity as $B_{\rm{exc}}=439~$T, which is close to the experimental value  for Mn inside NCs, 430~T \cite{norris}. When Mn is close to the surface, the emission becomes red shifted, and the  difference between $\sigma^+$ and $\sigma^-$ polarization is smaller than with Mn in the center.   For Mn in position II, the dark exciton $|\Downarrow \downarrow\rangle$ lies at lower energy than the bright ones, due to the geometrical relaxation induced by the electron-hole interaction. It seems that Mn atoms close to the surface would favor the emission through dark excitons.



\begin{figure}[!ht]
\centering
\includegraphics[scale=0.55]{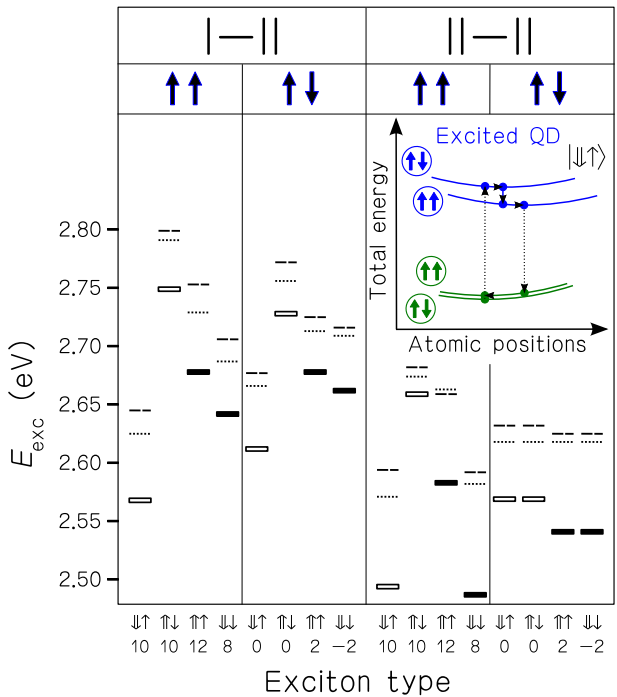}
\caption{\label{fig:fig3}(Color online)
The same as previous figure but the impurities positions are either I-II or II-II; and the magnetic alignment, either parallel or antiparallel. 
For bright excitons $\mu_{\rm{QD}}$ is either 10~$\mu_{B}$ or 0~$\mu_{B}$; for dark excitons $\mu_{\rm{QD}}$ is either 12~$\mu_{B}$, 8~$\mu_{B}$, 2~$\mu_{B}$ or -2~$\mu_{B}$.
The inset gives the total energy scheme for QDs doped with two Mn impurities in sites I-II. The Mn spins are noted with arrows within circles standing for the dots. The FM-AFM alignment difference for the excited states is about 10-50 meV, much larger than the ordering energies in the de-excited states ($\sim$1~meV). See that the ferromagnetic alignments 
of Mn atoms have lower excitation energies. }
\end{figure}

Next we look at NCs doped with two Mn impurities. The excitation energies $E_{\rm{exc}}$ are calculated for different positions and magnetic configurations of Mn spins, as plotted in Fig.~\ref{fig:fig3}. In case of bright-up excitons ($|\Downarrow\uparrow\rangle$), $E_{\rm{exc}}$ is  smaller when Mn impurities are placed in sites II-II than when placed in sites I-II, and this occurs regardless of their magnetic configurations. These differences can be understood in terms of the previous changes in the excitation spectra for Mn in positions I and II. When Mn impurities replace Cd atoms in sites II-II, the excitation energies corresponding to bright-up/down or dark-up/down excitons are degenerate because of symmetry (Fig.~\ref{fig:fig3}). 

Furthermore, {\it $E_{\rm{exc}}$ is smaller for parallel Mn spins than for antiparallel}. This magnetic effect is due to the effective exchange interactions between excitonic hole and Mn spins. The hole favors the parallel Mn alignment as in III-V semiconductors \cite{raebiger}. As a result of these interactions, the charge  distribution of the hole varies with the exciton type. The exciton energies after relaxing the geommetry departs farther from the fundamental gaps because positions II are involved.

We must now bring the finding concerning the Mn ferromagnetic alignment into contact with the experiments.

(a) When we excite the QD from the most stable configuration with antiparallel Mn magnetic moments, the system reaches a state with parallel Mn atoms. 
Spin flip lifetimes depend on the exchange coupling energy between Mn atoms in the system. In the excited state the large Mn-Mn exchange energy due to the exciton implies that the time for spin flips becomes smaller than the excitonic lifetimes \cite{besombes}. As the Mn moments prefer energetically to be aligned, they flip and align parallel. In fact, the formation of EMPs has already been observed in experiments,\cite{kuroda} which is related to excitons with long lifetimes and spin stability. After emission the magnetic state remains parallel for long time as the Mn-Mn exchange energy is very small, in the order of a meV \cite{carlos}. The long duration of the de-excited parallel state is $\sim10-100~\mu$s\cite{govorov}. Thus, the (Cd,Mn)Te QDs could behave as  magnetic nanoswitches, switchable from $\mu^{\ast}_{\rm{QD}}=0~\mu_B$ to $\mu^{\ast}_{\rm{QD}}=10~\mu_B$ by optical excitation. 

(b) We see that the difference between $\sigma^+$ and $\sigma^-$ changes for the parallel Mn alignment as compared to the antiparallel. The exciton states that are responsible for the  $\sigma^+$ ($\sigma^-$) photons shift towards the red (blue), and they are relatively more (less) occupied. The larger splitting between both polarization states means that the polarization  degree $\sigma^+$  or $\sigma^-$ increases when a magnetic polaron is formed in the dot, in agreement with the experiments \cite{mackowski}. We also found that the maximum shift in excitation energies by flipping spins is about 50 meV, which can be observed, for instance, by changing the temperature \cite{kuroda}. These shift values are larger than spin flips in the de-excited states by an order of magnitude. In consequence we could get magnetic polarons at higher temperatures than the typical ordering temperature. 

(c) For the lower bright excitation $|\Downarrow\uparrow\rangle$ the Stokes shift $\Delta_{\rm{SS}}$ gives an idea of the number of Mn atoms in the dot. The Stokes shift  of the absorption-emission process is defined as \cite{puzder} $\Delta_{\rm{SS}}=E_2-E_1-(E_3-E_4)$ after the inset of Fig.~\ref{fig:fig2}. In case of a single impurity, $\Delta_{\rm{SS}}$ is smaller when Mn substitutes the central Cd atom (site I, $\Delta_{\rm{SS}}^{\rm{I}}=47$~meV) than when an off-center Cd (site II, $\Delta_{\rm{SS}}^{\rm{II}}=124$~meV), close to the surface. 
These calculated Stokes shift values are in the same order of magnitude  than those for semiconductor dots of Si with $\sim 2$ nm of diameter, $\sim 0.1$~eV \cite{puzder}.
When Mn spins are parallel and placed in sites I-II, $\Delta_{\rm{SS}}^{\rm{I-II}}=224$~meV, and when placed  in sites II-II, $\Delta_{\rm{SS}}^{\rm{II-II}}=253$~meV. When assuming a vertical absorption for the antiparallel ground states, we have to add the spin flip energy of about 10-50 meV and the Stokes shift of the antiparallel states to the previous parallel values. For the configurations with low energy dark-excitons, we have to consider also in the Stockes shift the difference between bright and dark excitons \cite{perez}. Therefore,  as compared with the single impurity case these total Stokes shifts are about three times larger. Anyhow we see that significant differences in $\Delta_{\rm{SS}}$ could be used to detect the presence of a second Mn within NCs.


In summary, we have studied the exciton states of lowest energy in QDs of (Cd,Mn)Te within density functional theory, and related to them we have calculated excitation energies and total energies. For NCs with a single Mn impurity the excitation energy is found to be larger when the Mn atom substitutes a Cd atom in the QD center than when the Cd atom is close to the surface. For Mn near to the surface we see also that dark excitons become energetically more stable. For NCs with two Mn impurities the excitation energy is smaller for the parallel Mn spins than for the antiparallels. This magnetic effect is related to the hole, which mediates the exchange between Mn spins favoring parallel alignments. Due to hole-mediated exchange, the excitation of ground-state QDs with antiparallel Mn spins flips the magnetic state to parallel in the excited configuration. As the magnetic configuration remains parallel after emission for long ($\sim10-100~\mu$s), it indicates that  spin-based operations with (Cd,Mn)Te QDs could be controlled by light.
These results might be extrapolable to larger dots taking into account that the
excitation energies must be  red shifted due to the gap reduction at
larger sizes.
Future studies include the role of an electric field in the coupling
between Mn atoms.

This  work  was  supported   by  the  Basque  Government  through  the
NANOMATERIALS   project   (IE05-151)   under   the   ETORTEK   Program
(iNanogune), Spanish  Ministerio de  Ciencia y Tecnolog\'ia  (MCyT) of
Spain(  Grant  Nos.   TEC2007-68065-C03-03 and Fis2007-66711-CO2-02,  and  MONACEM  project)  and
University of the Basque Country (Grant No. IT-366-07).  The computing
resources from  the Donostia  International Physics Center  (DIPC) and
the SGI-SGIkerUPV are gratefully acknowledged.
 A. K. Bhattacharjee is acknowledged for
comments  on the manuscript.   C. E.-A.  wants to  thank N.  Gonz\'alez for
support on the plotting routines during this work.



\end{document}